# High quality $Fe_{3-\delta}O_4$/InAs hybrid structure for electrical spin injection.

Marhoun Ferhat[1*] and Kanji Yoh[1,2]

[1]CREST-JST, Kawaguchi, Saitama 332-0012 Japan
[2]Research Center for Integrated Quantum Electronics, Hokkaido University, Sapporo, 060-8628 Japa

(dated: 25 October 2006)

Single Crystalline $Fe_{3-\delta}O_4$ ($0 \leq \delta \leq 0.33$) films have been epitaxially grown on InAs (001) substrates by molecular beam epitaxy using $O_2$ as source of active oxygen. Under optimum growth conditions *in-situ* real time reflection high-energy electron diffraction patterns along with ex-situ atomic force microscopy indicated the (001) $Fe_{3-\delta}O_4$ to be grown under step-flow-growth mode with a characteristic $(\sqrt{2}\times\sqrt{2})R45$ surface reconstruction. X-ray photoelectron spectroscopy demonstrate the possibility to obtain iron oxides with compositions ranging from $Fe_3O_4$ to $\gamma$-$Fe_2O_3$. Superconducting quantum interference device magnetometer at 300K shows well behaved magnetic properties giving therefore credibility to the promise of iron based oxides for spintronic applications.

During the last decade there has been a considerable increase of studies on hybrid devices combining both semiconductors and magnetic materials. This is motivated by their possible applications in the nascent field of spintronics, where the electron spin degree of freedom is believed to be a source of a rich array of new physical phenomena. Different approaches have been adopted for the selection of adequate spin injectors including ferromagnetic metals,[1,2] dilute magnetic semiconductors[3] and Heusler alloys.[4] While magnetic semiconductors present the formidable task of increasing their Curie temperature, both ferromagnetic metal and Heusler alloys suffer from structural and interfacial problems during their growth. Quite surprisingly, there are little reports on iron oxides based half metals, which exhibit a large polarization at the Fermi level making them very attractive for spin injection into semiconductors.

$Fe_3O_4$ or magnetite belongs to a large family of iron based oxides commonly called ferrites. Our choice for this material as possible efficient spin injector into semiconductors stems from its physical properties, which are attractive in many respects: i) It has a half metallic character with large spin polarization at the Fermi level,[5,6] ii) It possess an electrical resistivity of the same order of magnitude as semiconductors making the conductivity mismatch problem less severe than in the case of metals iii) It has a Curie temperature of about 850K well above room temperature. On the other hand, InAs substrate was used obviously for its high room temperature electron mobility and more importantly it's large and tunable spin-orbit coupling strength (small ratio between the Rashba and Dresselhauss terms) important for spin-based devices.[7]

Although the growth of iron oxides on metallic and oxides based substrates is a very well documented subject[8], there are only scarce reports on the growth of iron oxides on semiconducting substrates.

The purpose of the present letter is to extend the only work available[9] on the growth of $Fe_3O_4$ on InAs with special emphasis on the growth mechanism and crystal quality using molecular beam epitaxy (MBE) technique.

Combining real time in-situ reflection high-energy electron diffraction (RHEED) and ex-situ atomic force microscopy (AFM) we demonstrate the possibility to obtain high Quality $Fe_3O_4$ (001) with good magnetic properties.

Epiready p-doped InAs (001) wafers were used as received. The samples, with typical size of about 7×10mm$^2$, were loaded in the solid source MBE system and the oxide layer was removed by annealing at around 510°C under an $As_4$ pressure of 1×10$^{-5}$ Torr. A 500-nm-thick InAs buffer layer was grown at 450°C using a beam-equivalent-pressure ratio of $As_4$ to In of about 12. This ensured the growth of high quality InAs buffer exhibiting a sharp and well ordered (2×4) surface structures termination.

Immediately after, the samples were transferred through an ultra high vacuum module, to an attached MBE chamber dedicated to the growth of iron oxides. The oxide MBE chamber is equipped with three metal beam sources, where a contact-less Fe rods were used to insure minimum contamination. The chamber was modified to include a high purity molecular oxygen reservoir with a precise variable leak valve for the control of the Oxygen partial pressure. An optimum growth temperature of 300°C with an oxygen partial pressure of 7.5×10$^{-7}$ and 4×10$^{-6}$ Torr were found to yield $Fe_3O_4$ and $\gamma$-$Fe_2O_3$ respectively. The calibration of the growth temperature was performed by carefully monitoring the (2×4)→(4×2) InAs surface structure transition (about 410°C) as well as the melting point of In (157°C). In all case the growth rate were fixed to about 10nm/h.

The RHEED patterns of the InAs (001) along the [110] and [-110] azimuths together with those of 20nm-thick-$Fe_3O_4$ along the [110] and [100] azimuths of InAs (001) are shown in Figure1. The RHEED patterns [Fig.1(c) and 1(d)] are similar to those observed[10,11] in epitaxial $Fe_3O_4$/MgO (001) but rotated 45° relatively to the [100] azimuth of InAs (001). The epitaxial relationship is therefore [110] $Fe_3O_4$ (001)//[100] InAs (001) with a clear $(\sqrt{2}\times\sqrt{2})R45$ surface structures.

The lattice constant along [110] was found to be the same as that of InAs ($a_{[110]}$ = 8.57Å) indicating the full pseudomorphic nature of the films. This explains well the 45° rotation of $Fe_3O_4$ ($a_{[100]}$ =8.40Å) relatively to bare InAs, where the lattice mismatch is reduced to only 1.9%. A similar situation was recently encountered in $Fe_3O_4$/GaAs[12] although the large lattice mismatch of 5% leads probably to films quality inferior to the present case. However, even a lattice mismatch of 1.9% is not expected to yield the sharp and streaky patterns we observe in Fig. 1(c) and 1(d), this is because such a highly tensile strain should induce 3D growth mode, which manifest as spots in the RHEED patterns.

*Electronic address: ferhat@eng.hokudai.ac.jp

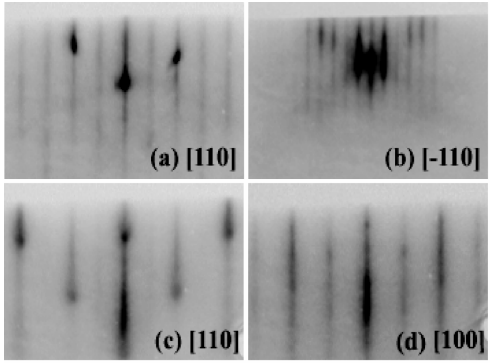

FIG. 1. RHEED patterns of bare (001) InAs along (a) the <110> and (b) <-110> just before the growth and 20nm-thick $Fe_3O_4$ along (c) the <110> and (d) the <100> azimuths. All azimuths are refereed to the InAs substrate.

Figure 2(a) shows an AFM topographic of an as-grown $Fe_3O_4$ films with macroscopic step edges (up to 10nm height) separated by flat terraces (up to 2µm wide) running along the [-110] direction of the InAs substrate. Close inspection of the terraces shows them to be atomically flat with rectangular features [fig 2(b)] probably reflecting domains with different surface termination.[13]

Both our AFM images together with the absence of RHEED oscillations during the growth are strong evidence of a step-flow-growth mode in the present case. In order to explain the occurrence of this particular growth mode, recall that in contrast to GaAs (001) case, the (2×4)→(4×2) phase transition in InAs (001) is of first-order nature owing to the strong lateral interaction of As-surface unit in the (2×4) structure of InAs (001).[14,15] Consequently, the As-rich surface structures is well ordered and highly uniform. This is particularly true in our case as can be clearly seen in the RHEED patterns of InAs [Fig. 1(a) and 1(b)] just before the growth. The large size of our sample holder however, required about 30min to stabilize the substrate temperature. This relatively prolonged annealing at 300°C leads to preferential As-desorption with the creation of monomolecular steps running along the [-110] (direction of the As dimmers rows) as in the case of reference 15 and are the origin of the step-flow-growth mode observed. However, the steps height in Fig. 2(a) is not monomolecular and some terraces are over 1µm wide, which is evidence of step-bunching instability. Two mechanisms are qualitatively relevant to our experimental results. The first mechanism proposed by Tersoff et al[16] has its origin in the large elastic strain of our pseudomorphic films (1.9% mismatch), the elastic strain relaxation at steps produce long-ranged attractive interaction between steps and therefore steps-bunching. The second mechanism proposed by Kandel and Weeks[17] is related to the surface reconstruction, the presence of differently reconstructed area on the terraces, May impedes the motion of step thus decreasing its velocity and leading to step-bunching. The rectangular features observed in Fig. 2(b) very likely represent areas with different surface termination, thus kinetically leading to steps-bunching instability. The existence of some terraces extending up to about 2µm wide suggests the combined action of both strain and surface reconstruction mechanisms during the growth.

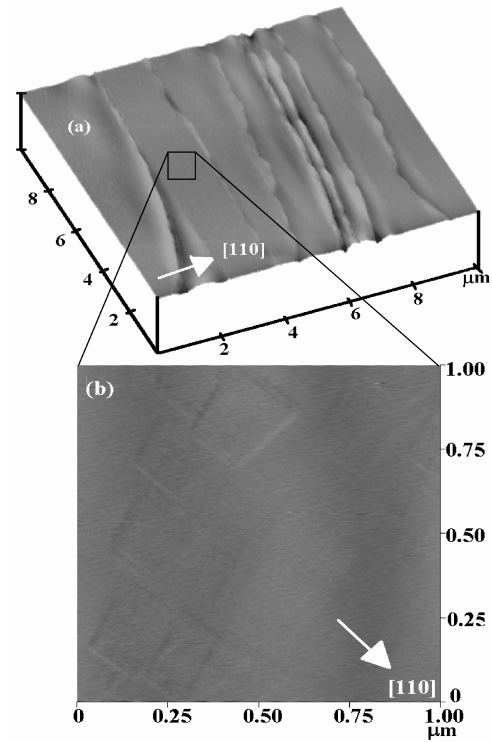

FIG. 2. Typical AFM images of a 20nm-thick $Fe_3O_4$/InAs (a) 3D topography showing stepped surface with large terraces (b) enhanced view of a terraces showing rectangular features attributed to domains with different surface termination.

Although $Fe_3O_4$ has a rather similar crystal structure to that of $\gamma$-$Fe_2O_3$, in typical XPS measurements, a distinctive satellite structures due to charge transfer screening appears in the Fe 2p core level spectrum of $\gamma$-$Fe_2O_3$ but not in that of magnetite. Figure 3(a) and 3(b) shows the high-resolution Fe 2p XPS spectra (Mg K$\alpha$) of 20nm-thick iron oxide films grown at the same substrate temperature of 300°C but under an $O_2$ partial pressure of $7.5 \times 10^{-7}$ and $4 \times 10^{-6}$ Torr respectively. All binding energies were corrected of charging effect by setting the O 1s at 530.1eV. Clearly, figure 3(b) exhibit a "shake-up" satellite at about 719.7eV, which is characteristic of $Fe^{3+}$ ions indicating therefore the formation of $\gamma$-$Fe_2O_3$. Since $Fe_3O_4$ contains both $Fe^{2+}$ and $Fe^{3+}$ ions, their mutual contribution to the Fe 2p XPS spectra between the main spin-orbital (Fe2$p_{1/2}$ and Fe2$p_{3/2}$) give rise to a vanishing and unresolved component. The absence of the "shack-up" satellite in Fig. 3(a) can be considered as evidence of near stiochiometric $Fe_3O_4$. This is somehow expected since a large $O_2$ partial pressure was required to obtain

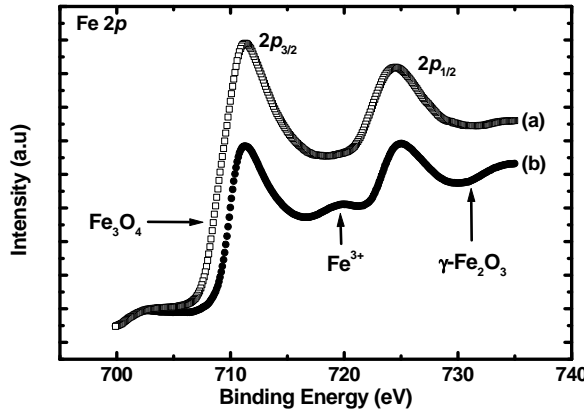

FIG. 3. Fe 2$p$ high-resolution XPS spectra for 20nm-thick Fe$_{3-\delta}$O$_4$/InAs grown at 300°C under an oxygen partial pressure of (a) $7.5\times10^{-7}$ Torr and (b) $4\times10^{-6}$ Torr. The compositions obtained are indicated in the figure.

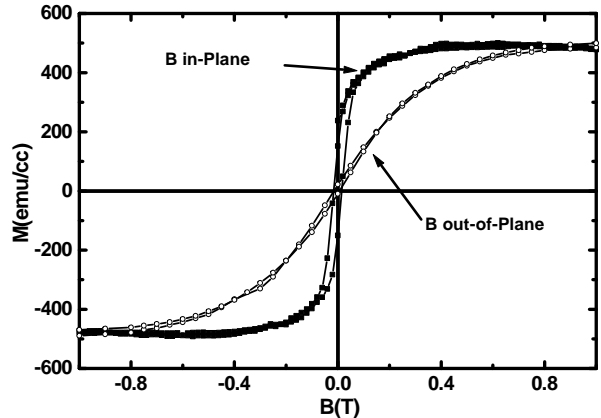

FIG. 4. Room temperature hysteresis loop obtained for 20nm-thick Fe$_3$O$_4$/InAs with magnetic field applied both parallel and perpendicular to film plane directions.

$\gamma$-Fe$_2$O$_3$, which can be considered as an oxidized form of Fe$_3$O$_4$.

Strictly speaking, the growth of Fe$_3$O$_4$ yield in general non-stoichiometric magnetite, which should be written as Fe$_{3-\delta}$O$_4$ with $\delta=0$ for pure magnetite and $\delta=0.33$ for maghemite or $\gamma$-Fe$_2$O$_3$. Quantification of the vacancy parameter $\delta$ is quite challenging but a close comparison of the shape of the XPS spectra in reference 18 (Fig. 9) to those depicted in Fig. 3 indicates that deviation from stoichiometry if present is likely to be very small in our films.

Figure 4 shows the hysteresis loops measured at 300K for the optimized Fe$_3$O$_4$ film. It exhibits an easy in-plane magnetization [Fig. 4(a)] due to the shape anisotropy with a coercive field of about 0.022 T and a saturation magnetization of about 483 emu/cc in close agreement with the value reported for bulk magnetite.[19] The in-plane saturation was achieved at a field of about 0.5T while a field of about 1T was necessary for the out-of- plane direction. This is in contrast with the previous results,[20,21] where the presence of antiphase boundaries (APBs) leads to unsaturated magnetization at field as large as 7T. A plausible explanation to this difference should be sought in the growth mechanism. In a typical 2D layer-by-layer mode the coalescence of nuclei during the growth is likely to form APBs extending along all the film thickness.[22] In step-flow-growth mode however, the nucleation occurs preferentially at the steps edges and the growth extend laterally leading to reduced APBs density. Perhaps the present situation is similar to the case of GaAs on Ge,[23] where APBs free films were obtained by appropriate surface treatment.

In summary, high quality (001) Fe$_{3-\delta}$O$_4$ films were epitaxially grown on (001) InAs substrate using molecular Beam epitaxy. The Fe 2$p$ XPS core level spectra demonstrate the possibility to grow in controlled manner both Fe$_3$O$_4$ and $\gamma$-Fe$_2$O$_3$ with small deviation from stoichiometry. The magnetic properties of the films shows a bulk-like behavior tentatively explained by reduced APBs owing to the step-flow-growth mode observed. The overall results gives credibility to the promise of iron based oxides for spintronic applications.

The authors are grateful to A. Subagyo, K. Sueoka and Y.suda for useful discussions. This work was partially supported by the Japanese Ministry of Education, Culture, Sports, Science and Technology.